\documentclass[aps,prl,twocolumn,superscriptaddress]{revtex4}  

\usepackage{amsmath}
\usepackage{amssymb}
\usepackage{graphicx}
\usepackage{wasysym}
\usepackage{color}
\usepackage{cancel}
\usepackage[colorlinks,linkcolor=blue,citecolor=blue,anchorcolor=green]{hyperref}
\include{papersdefinitions}
\def\ket#1{\left|#1\right>}
\newcommand{\be}{\begin{equation}}
\newcommand{\ee}{\end{equation}}
\newcommand{\bea}{\begin{eqnarray}}
\newcommand{\eea}{\end{eqnarray}}
\newcommand{\rmd}{\mathrm{d}}

\begin{document}

\title{Multi-pulse addressing of a Raman quantum memory: configurable beam splitting and efficient readout.}

\author{K.~F.~Reim}
\email[]{reimk@phys.ethz.ch}
\affiliation{Clarendon Laboratory, University of Oxford, Parks Road, Oxford OX1 3PU, UK}
\affiliation{Department of Physics, ETH Z\"urich, CH-8093 Z\"urich, Switzerland}

\author{J.~Nunn}
\affiliation{Clarendon Laboratory, University of Oxford, Parks Road, Oxford OX1 3PU, UK}

\author{X.-M.~Jin}
\affiliation{Clarendon Laboratory, University of Oxford, Parks Road, Oxford OX1 3PU, UK}
\affiliation{Centre for Quantum Technologies, National University of Singapore, 117543, Singapore}

\author{P.~S.~Michelberger}
\affiliation{Clarendon Laboratory, University of Oxford, Parks Road, Oxford OX1 3PU, UK}

\author{T.~F.~M.~Champion}
\affiliation{Clarendon Laboratory, University of Oxford, Parks Road, Oxford OX1 3PU, UK}

\author{D.~G.~England}
\affiliation{Clarendon Laboratory, University of Oxford, Parks Road, Oxford OX1 3PU, UK}

\author{K.C.~Lee}
\affiliation{Clarendon Laboratory, University of Oxford, Parks Road, Oxford OX1 3PU, UK}

\author{N.~K.~Langford}
\affiliation{Clarendon Laboratory, University of Oxford, Parks Road, Oxford OX1 3PU, UK}
\affiliation{Department of Physics, Royal Holloway, University of London, Egham Hill, Egham TW20 0EX, United Kingdom}

\author{I.~A.~Walmsley}
\affiliation{Clarendon Laboratory, University of Oxford, Parks Road, Oxford OX1 3PU, UK}

\date{\today}

\begin{abstract}
Quantum memories are vital to the scalability of photonic quantum information processing (PQIP), since the storage of photons enables repeat-until-success strategies. On the other hand the key element of all PQIP architectures is the beam splitter, which allows to coherently couple optical modes. Here we show how to combine these crucial functionalities by addressing a Raman quantum memory with multiple control pulses. The result is a coherent optical storage device with an extremely large time-bandwidth product, that functions as an array of dynamically configurable beam splitters, and that can be read out with arbitrarily high efficiency. Networks of such devices would allow fully scalable PQIP, with applications in quantum computation, long-distance quantum communications and quantum metrology.

\end{abstract}

\maketitle

Quantum information processing promises radical new technologies, such as provably secure communications \cite{Gisin_2002_RMP_Quantum-cryptography}, enhanced sensors \cite{Thomas-Peter2011PRL_Real-World-Quantum-Sensors:-Ev}, and super-fast computing \cite{Harrow:2009fk}. Photonics offers a powerful platform for quantum processing, with interactions between optical modes mediated by beam splitters. However, logical operations are conditioned on measurements \cite{Knill_2001_N_A-scheme-for-efficient-quantum}, which means all PQIP schemes are probabilistic, requiring optical quantum memories \cite{Simon2010EPJ_Quantum-memories---A-review-ba} to synchronize elements. In this paper, we demonstrate that addressing a Raman quantum memory \cite{Reim_2010_P_Towards-high-speed-optical-qua} with a train of control pulses produces an interaction that is formally identical to a dynamically configurable network of beam splitters. Recent results have established the coherence of this interaction for the case of photon-echo Raman memories \cite{Campbell2011A_Coherent-Time-Delayed-Atom-Lig,Sparkes:2012uq}. In the present work, we use a broadband Raman memory protocol \cite{Reim2011PRL_Single-Photon-Level-Quantum-Me} with a time-bandwidth product --- the ratio of storage time to pulse duration --- in excess of 1000. In this regime, multiple independent readout pulses can be used to trigger a train of read events. This allows to distribute photons over many temporal modes with controllable amplitudes, and, furthermore, to extract excitations stored in the memory with arbitrarily high efficiency. The combination of storage and programmable coherent coupling in a single device makes the Raman memory a universal primitive component for PQIP.

As an example, consider the task of \emph{photon subtraction} \cite{Kitagawa2006PR_Entanglement-evaluation-of-non}, which is a canonical non-Gaussian operation in continuous-variables (CV) quantum information protocols \cite{Braunstein_2005_RMP_Quantum-information-with-conti}. Traditionally, one half of a two-mode squeezed state is directed through a highly transmissive beam splitter. When a photon is detected at the reflected port, the transmitted joint state is no longer Gaussian, and is more entangled. However, the detection probability is small, so that many squeezed states must be prepared to achieve success. On the other hand, if one half of a squeezed state is stored in a Raman memory, a train of weak read pulses --- corresponding to multiple beam splitters --- allows to repeat the protocol until success (that is, until the emission of a photon from the memory is detected). This application permits a doubly-exponential enhancement in the resource-efficiency of CV entanglement distillation \cite{Datta:2012fk}.

\begin{figure}[h]
\begin{center}
\includegraphics[width=\columnwidth]{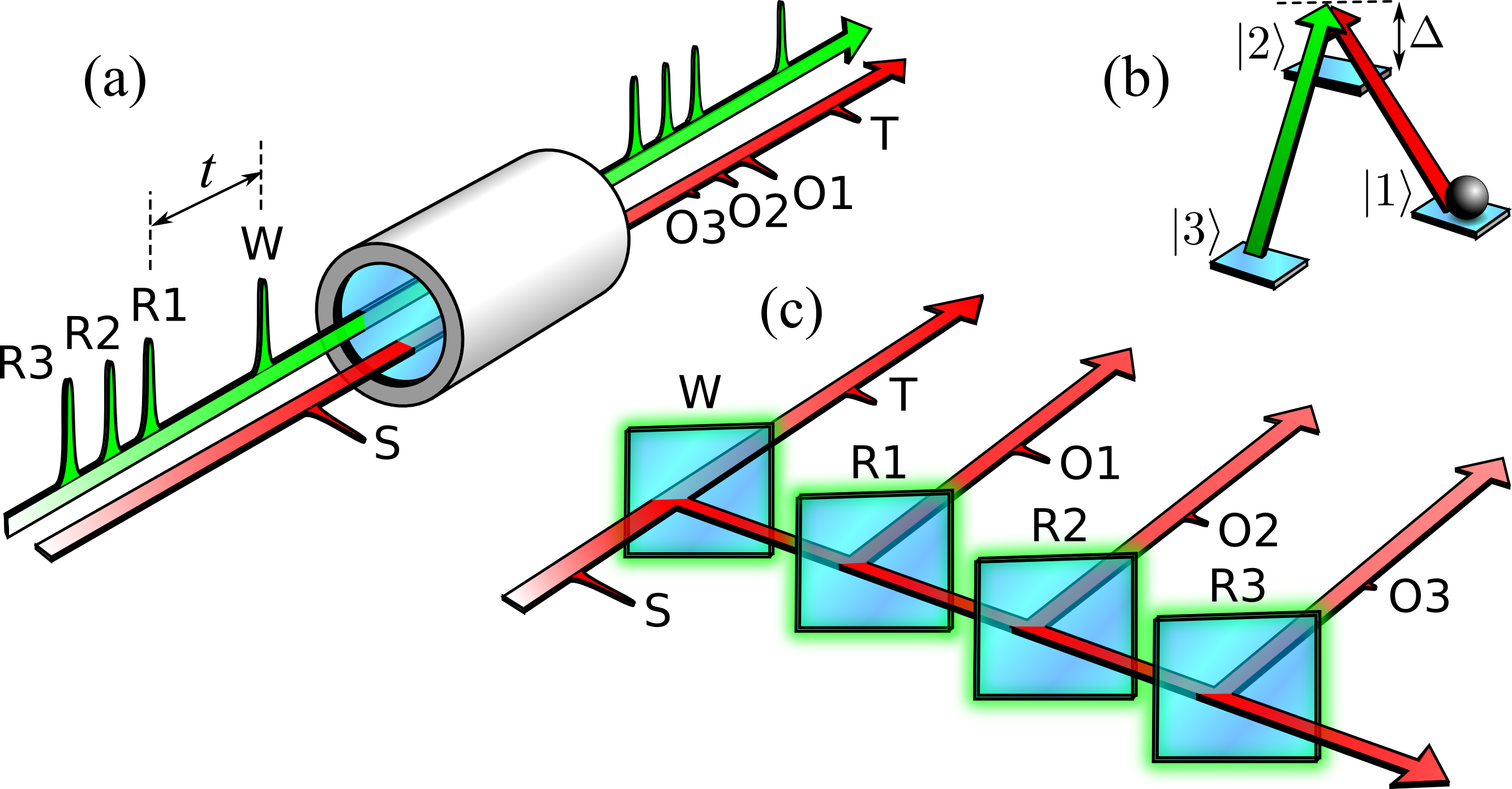}
\caption{(a) Raman memory: an incident signal S is directed through a Cs vapour cell along with an orthogonally polarized `write' control pulse W, followed by a series of `read' pulses R1--R3. The unstored portion T of the signal is transmitted through the memory, while the retrieval pulses extract outputs O1--O3. (b) Lambda-type level structure of the Cs atoms: atoms are prepared in $\ket{1}$ (grey sphere). The signal (red) is blue-detuned from the excited manifold $\ket{2}$ and is coupled to the storage state $\ket{3}$ via the Raman-resonant control (green). (c) Beam splitter network equivalent to (a).}
\label{fig:memory_and_beam_splitters}
\end{center}
\end{figure}

Storage in atomic ensembles allows strong light-matter coupling without the need for high-finesse cavities \cite{Specht2011N_A-single-atom-quantum-memory,Kuhn_2002_PRL_Deterministic-Single-Photon-So}; furthermore multiple spatio-temporal modes \cite{Usmani:2010lr,Nunn_2008_PRL_Multimode-Memories-in-Atomic-E,Grodecka-Grad:2011kx}, or multi-photon (\emph{i.e.} CV) states \cite{Jensen_2011_P_Quantum-memory-for-entangled-c,Appel_2008_PRL_Quantum-Memory-for-Squeezed-Li} can be stored. It has recently been shown that off-resonant Raman memories can also operate with quantum-limited noise in room-temperature vapours \cite{Hosseini2011C_High-efficiency-coherent-optic,Reim2011PRL_Single-Photon-Level-Quantum-Me}, since collisional fluorescence \cite{Manz_2007_PR_Collisional-decoherence-during,Hsu_2006_PRL_Quantum-Study-of-Information-D} is eliminated. This off-resonant suppression of spontaneous emission makes the atom-light interaction \emph{unitary}, so that via the Bloch-Messiah reduction \cite{Braunstein:2005vn,Wasilewski:2006ys} it is formally isomorphic to a sum of independent beam splitter transformations \cite{Nunn_2007_PRAMOP_Mapping-broadband-single-photo,Nunn_2006_Q_Modematching-an-optical-quantu}. Essentially, Raman memories behave like beam splitters because the `failure mode' is transmission: light that is not stored exits the memory unattenuated, since there is no resonant absorption. Similarly a stored excitation --- a spin wave --- that is not retrieved simply remains in the memory, since no population is transferred to the excited state. Our far off-resonant memory has a $>$GHz acceptance bandwidth, meaning that we are able to store, and subsequently address the memory with, many short optical pulses within the $\sim\mu$s decoherence time of the memory. In this paper we show how to utilize this large time-bandwidth product ($\geq 1000$) to implement complex and configurable linear optical networks by temporal multiplexing with a train of readout control pulses (as represented in parts (a) and (c) of Fig.~\ref{fig:memory_and_beam_splitters}). We demonstrate that a stored excitation can be completely retrieved, with efficiency arbitrarily close to unity, with such a train of pulses. Finally we verify that coherence is preserved across temporal modes by interfering signals retrieved from two independent memories.

Our quantum memory is based on Raman absorption in a warm vapour cell (62~$^\circ$C, 7~cm long) containing cesium and 20~Torr of Ne buffer gas (shown schematically in part (a) of Fig.~\ref{fig:memory_and_beam_splitters}). The Cs atoms are prepared in their ground states $\ket{1}$ (6$^2S_{1/2}$ F=4) via optical pumping with a diode laser at 852~nm (150~$\mu$W, beam waist 350~$\mu$m). The signal field to be stored (pulse duration 300~ps) is tuned away from the atomic resonance $\ket{2}$ (6$^2P_{3/2}$) by a large detuning $\Delta=15$~GHz but is coupled to the long-lived 9.2~GHz hyperfine-shifted storage state $\ket{3}$ (6$^2S_{1/2}$ F=3) by a bright control pulse (also 300~ps), which is tuned into two-photon resonance with the signal (see part (b) of Fig.~\ref{fig:memory_and_beam_splitters}). After some delay $t$ chosen by the user, subsequent control pulses convert the spin wave back into optical signals.

We model the effect of applying multiple readout pulses using a linearized, adiabatic theory of Raman storage developed previously \cite{Reim_2010_P_Towards-high-speed-optical-qua,Nunn_2007_PRAMOP_Mapping-broadband-single-photo}. Given an initial input pulse with profile $A_\mathrm{in}(\tau)$, the spatial distribution of excitations in the memory is given by the spin-wave amplitude
\begin{equation}
\label{Bz}
B_\mathrm{out}(z) = \int K_\mathrm{s}[\Omega_\mathrm{s}(\tau);\tau,z] A_\mathrm{in}(\tau)\,\rmd \tau,
\end{equation}
where $K_\mathrm{s}$ is an integral kernel involving a Bessel function \cite{Reim_2010_P_Towards-high-speed-optical-qua,Nunn_2007_PRAMOP_Mapping-broadband-single-photo}. The argument $\Omega_\mathrm{s}(\tau)$ is included to indicate the explicit dependence of $K_\mathrm{s}$ on the temporal shape of the control pulse used to store the signal field. For readout, the retrieved signal field is given by a similar formula,
\begin{equation}
\label{Aout}
A_\mathrm{out}(\tau) = \int K_\mathrm{r}[\Omega_\mathrm{r}(\tau);z,\tau] B_\mathrm{in}(z)\,\rmd z,
\end{equation}
where $K_\mathrm{r}$ is similar to $K_\mathrm{s}$, and where $\Omega_\mathrm{r}$ now describes the time-dependence of the control field used to retrieve the stored spin-wave. As shown in \cite{Nunn_2007_PRAMOP_Mapping-broadband-single-photo,Nunn_2006_Q_Modematching-an-optical-quantu}, singular value decomposition of these kernels provides a set of optical and spin wave modes that are independently coupled by effective beam splitter interactions, where the reflectivity is set by the energy in the control. In general we have that $B_\mathrm{in}(z) = \kappa B_\mathrm{out}(z)$, where $\kappa<1$ is a constant representing decay of the spin wave amplitude through decoherence over the storage time. Note that in the case of multi-pulse retrieval, $\Omega_\mathrm{r}$ has the form of a train of pulses. We experimentally determine the temporal profiles $A_\mathrm{in}$, $\Omega_\mathrm{s,r}$ by taking square roots of the traces measured using a fast photodiode (see below). This assumes flat phase for all pulses but autocorrelation measurements confirm that the laser output is transform-limited. 

\begin{figure}[h]
\begin{center}
\includegraphics[scale=0.42]{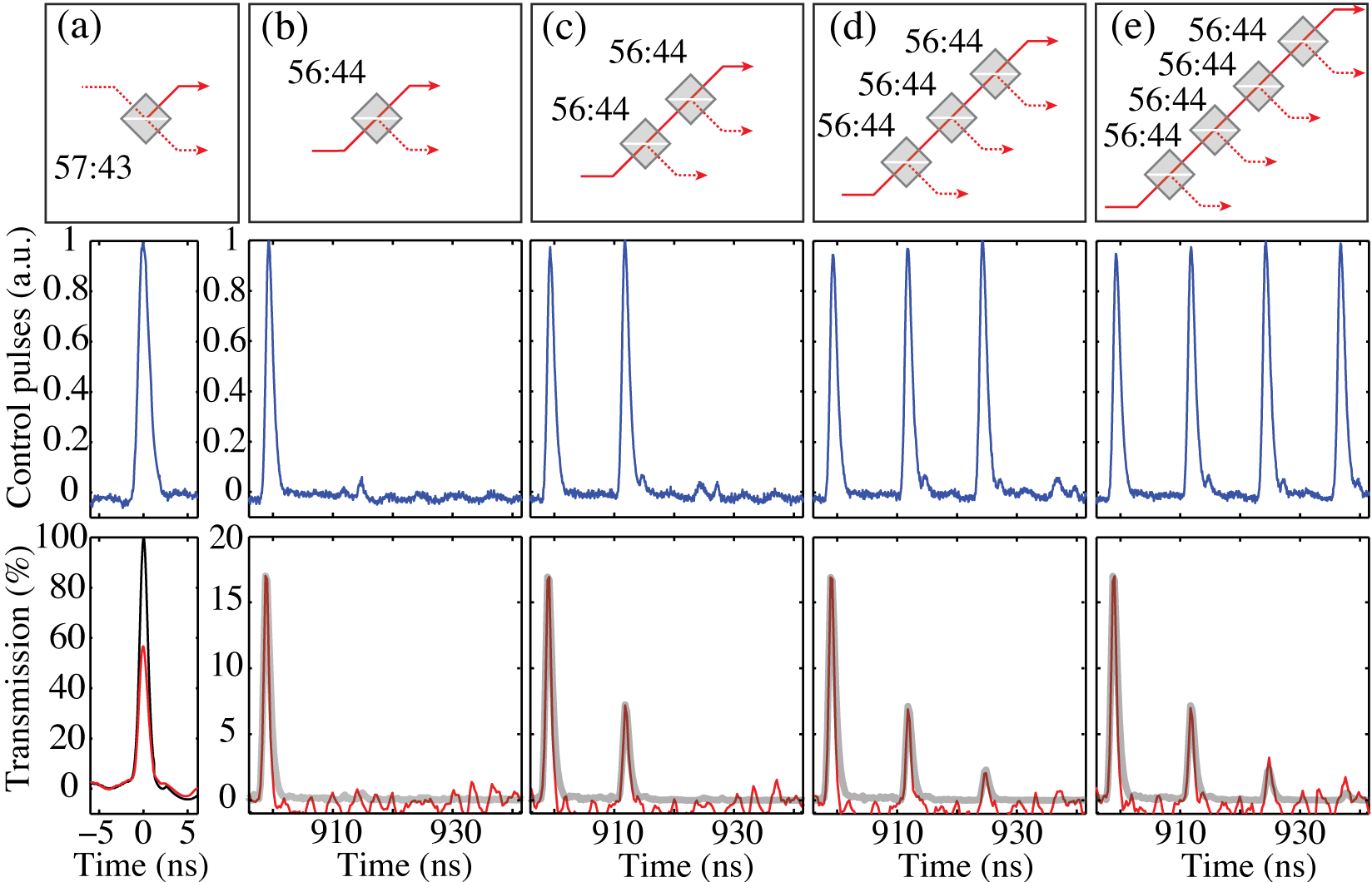}
\caption{Multiple retrieval. Blue lines: control field pulses. Black line: incident signal field. Red lines: transmitted / retrieved signal pulses. Thick grey lines: theoretically predicted retrieved signals.
\textbf{(a)} Storage event at $t=0$~ns. \textbf{(b)} Single retrieval at $t=899$~ns. \textbf{(c)} Double retrieval at $t=899$~ns and $t=911.5$~ns. \textbf{(d)} Triple retrieval at $t=899$~ns, $t=911.5$~ns and $t=924$~ns. \textbf{(d)} Quadruple retrieval at $t=899$~ns, $t=911.5$~ns, $t=924$~ns and $t=936.5$~ns. The last retrieved pulse is too weak, so that it vanishes in the noise. The storage efficiency is 43\%. The retrieval efficiency for each read event is 56.5\% leading to total readout efficiency of 17\% for the first read pulse, 7.5\% for the second read pulse, 3\% for the third read pulse and $\sim$1\% for the fourth pulse. The top panels show the analogous beam splitter networks.}
\label{fig:Multiple_retrieval}
\end{center}
\end{figure}

To implement multi-pulse readout, we use a Pockels cell (Quantum Technology Starfire 5DR) to select a series of consecutive pulses (energy $\sim$10~nJ) separated by $12.5$~ns from the 80~MHz pulse train of the laser (Newport Tsunami). We store a weak signal pulse (a coherent state containing around 1000 photons), frequency-shifted from the control by means of a 9.2~GHz electro-optic modulator (Newport~4851-M), and observe the transmitted and retrieved pulses on an amplified avalanche photo-diode with a ns response time (Thorlabs~APD210). Fig.~\ref{fig:Multiple_retrieval} shows the successive retrieval of between 1 and 4 pulses; the agreement with theory is excellent. The model predicts that each read pulse should extract roughly 57\% of the stored spin wave, and this explains the observed efficiencies correctly, after accounting for decoherence of the spin wave by $\sim$30\% ($\kappa\approx0.7$) over the 900~ns storage time, which is due mostly to the influence of stray magnetic fields \cite{Reim2011PRL_Single-Photon-Level-Quantum-Me}.

The combined retrieval efficiency of four read pulses is 95\%, which demonstrates how unit retrieval efficiency is rapidly approached with a multi-pulse read sequence. The efficiency of an ensemble memory is generally limited by the achievable resonant optical depth, whether or not the protocol operates on resonance \cite{Phillips_2008_PR_Optimal-light-storage-in-atomi,Gorshkov_2007_PRL_Universal-Approach-to-Optimal-,Amari:2010zr}, but here this number exceeds 1000 \cite{Reim2011PRL_Single-Photon-Level-Quantum-Me} and is not a limiting factor, which is why unit efficiency can be reached.

In general, partial storage is useful as a means to entangle optical and material modes. Also, non-deterministic storage can be heralded by post-selecting those events where no light is transmitted through the memory. By contrast, inefficient retrieval from the memory is always detrimental, since all protocols involving memories eventually require that stored excitations are read out. Therefore the ability to completely retrieve the stored state, with arbitrarily high efficiency, using multi-pulse retrieval, is an important capability of the Raman memory.

Besides efficient retrieval, adjustment of the readout sequence enables control of the distribution of the retrieved signal over the readout modes. As discussed above, the theoretical model of the Raman interaction has the form of a collection of independent beam splitter transformations with effective reflectivities determined by the control field \cite{Nunn_2007_PRAMOP_Mapping-broadband-single-photo,Nunn_2006_Q_Modematching-an-optical-quantu}. To further test the assertion that our memory can be described by this model, and to demonstrate the configurability of the beam splitter reflectivities, we varied the intensity distribution of the read control pulses by adjusting the Pockels cell timing.

\begin{figure}[h]
\begin{center}
\includegraphics[width=\columnwidth]{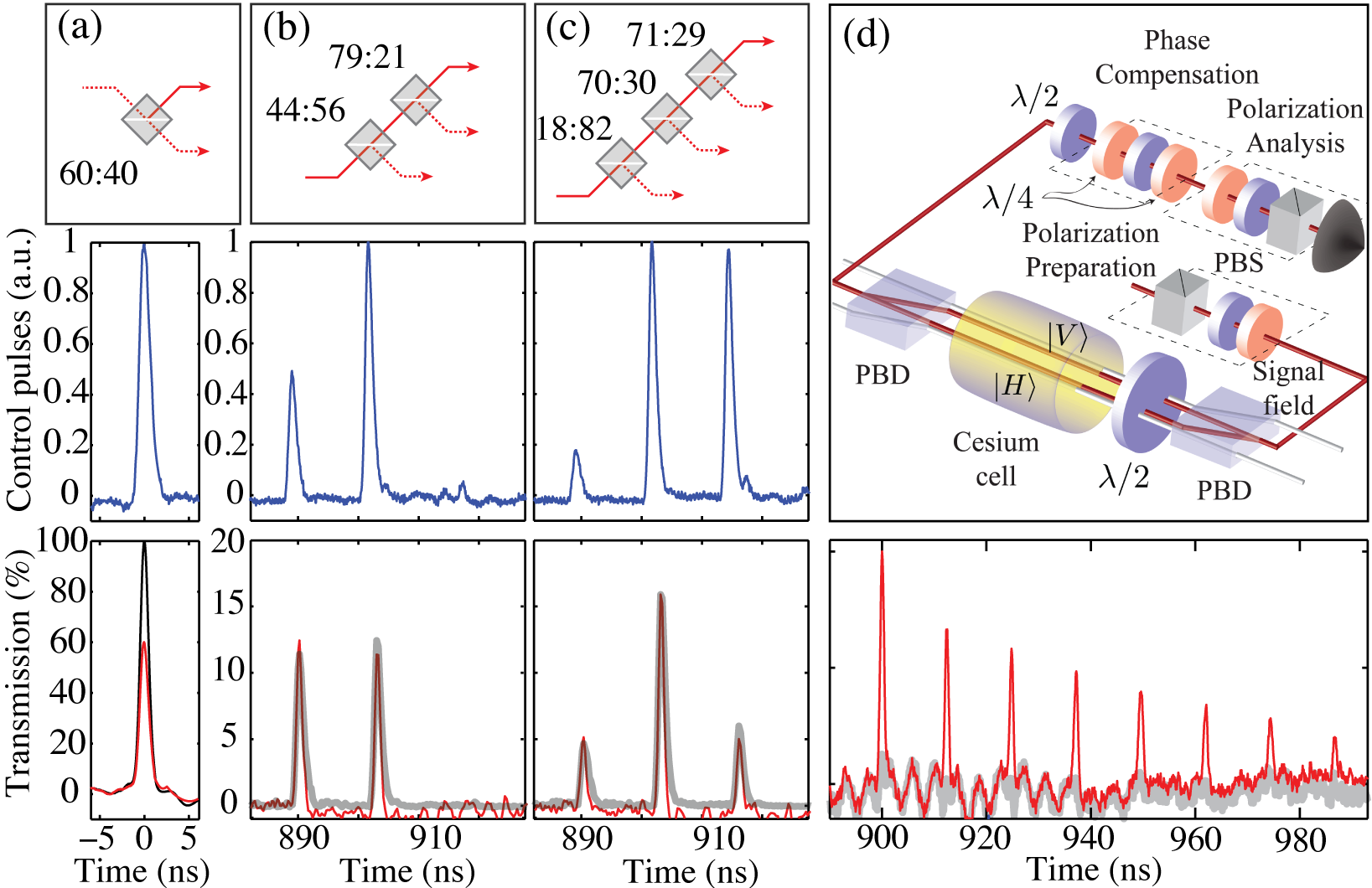}
\caption[Shaping and coherence of the readout]{Shaping and coherence of the readout. Colour coding in (a), (b), (c) is the same as in Fig.~\ref{fig:Multiple_retrieval}.  \textbf{(a)} Storage at $t=0$ ns. \textbf{(b)} Balanced readout: retrieval with two increasingly strong control field pulses starting at $t= 888$~ns. \textbf{(c)} Pyramidal readout: retrieval with three increasingly strong control field pulses starting at $t= 888$~ns. The storage efficiency is 40\%. The retrieval efficiency for the balanced readout is 12.4\% each, and the retrieval efficiency for the pyramidal readout is 5\% for the first, 16\% for the second and 5\% for the third retrieved signal. Top panel shows analogous beam splitter networks. \textbf{(d)} Coherence of multi-pulse readout demonstrated by interference between two pulse trains retrieved from independent memories. Top panel: experimental layout. PBD: polarizing beam displacer; PBS: polarizing beam splitter. $\lambda/2$; $\lambda/4$: half- and quarter-wave plates. The `phase compensation' waveplates correct for unavoidable birefringence in the interferometer. Bottom panel: constructive and destructive interference across the entire multi-pulse retrieval: the incident signal is prepared with diagonal polarisation and is split equally between the two memories. We analyze the retrieved signals in either diagonal (red) or anti-diagonal (grey) polarizations.}
\label{fig:Multiple_retrieval2}
\end{center}
\end{figure}

The close agreement between experiment and theory in Fig.~\ref{fig:Multiple_retrieval2} confirms that the interaction can indeed be described as a beam splitter interaction between optical and material modes. These data also demonstrate the shaping of a stored signal pulse into multiple retrieved signals in different time bins \cite{Novikova_2008_PRAMOP_Optimal-light-storage-with-ful}, which approach offers a straightforward path to generating time-bin entanglement. For instance, storage of a single photon, followed by partial retrieval using three read pulses creates a time-bin entangled $W$-state \cite{Dur:2000ve}, in which the retrieved signal photon is distributed over three temporally distinct modes.

Finally, we tested the coherence of the interaction across multiple time bins by interfering the output of two independent memories. To do this, we inserted the vapour cell into a passively stable Jamin-Lebedev interferometer (a polarization Mach-Zehnder), and illuminated each arm with optical pumping and control pulses, creating a pair of parallel, independent memories, whose output could be interfered by observing signals in the diagonal polarization basis. Part (d) of Fig.~\ref{fig:Multiple_retrieval2} shows that we achieve near-complete destructive interference across an eight-pulse retrieval, which is possible only if the interaction is coherent across all time-bins.

In summary, we have demonstrated that a far off-resonant Raman memory can serve both as a high bandwidth storage device and as a programmable and actively tuneable beam splitter array. We used a train of control pulses to distribute a stored excitation among multiple time bins with adjustable amplitudes; we showed that near-perfect retrieval efficiency can be achieved with the use of just four read pulses, and we demonstrated that phase coherence is preserved across all time bins. Devices based on this interaction could form the core of a scalable platform for PQIP, enabling fast and reconfigurable quantum networking with in-built storage.

\section*{Acknowledgements}
This work was supported by EPSRC through the projects QIP IRC (GR/S82716/01) and EP/C51933/01 and the European Community's Seventh Framework
Programme FP7/2007-2013 under grant agreement n$^\circ$ 248095 for project Q-ESSENCE, and the Royal Society, and the AFOSR through the EOARD. XMJ acknowledges support from NSFC(No. 11004183) and CPSF (No.201003327). KFR was supported by the Marie-Curie ITN EMALI and PM and TFM were supported by the ITN FASTQUAST. The authors acknowledge helpful discussions with M.~Barbieri.


\end{document}